# General Theory of Relational Primacy

Ngueuleweu Tiwang Gildas

Laboratory of Applied Economics and Management Sciences,
Department of Agribusiness, Faculty of Economics and Management,
University of Dschang, Cameroon

gildas.ngueuleweu@ univ-dschang.org

**Abstract.** This paper presents the General Theory of Relational Primacy (GTRP), a conceptual and formal framework for understanding stability, crisis, and transition in complex systems. The central thesis is that systems are not defined by their variables but by the relations that organize them. Variables are merely late manifestations of deep relational mechanisms. Crises arise from the silent reorganization of these relations, not from changes in the variables themselves. We formalize three axioms (relational primacy, antecedence of relations, and universality of coupling) and derive from them a complete hypothetico-deductive framework (hypotheses $H_1$–$H_5$, validity conditions $C_1$–$C_6$, properties $P_1$–$P_5$ of the observable $\beta$). We prove a foundational theorem establishing that every complex system possesses an irreducible relational level, introduce an invariant of the theory analogous to the conservation of energy, and propose universal definitions of crisis, relational causality, and relational information. The parameter $\beta(t) = d \log y / d \log x$, estimated via a time-varying Kalman filter in the log-log space, measures this reorganization with an $O(\delta\theta)$ sensitivity, one order of magnitude faster than classical $O(\delta\theta^2)$ signals. The empirical evidence (NASA AIRS data, 2002–2026) and validations on six simulated systems confirm that $\beta$ precedes classical early warning signals by 14 to 24 months and 39 to 153 time steps. The theory is designed to be universal, applicable to climate, biology, economics, artificial intelligence, and sovereignty of systems.

# 1. Ontological foundations

## 1.1. Why the sciences err by observing variables

### 1.1.1. The displacement in the history of sciences

The history of sciences is marked by a series of fundamental displacements of the object of study. Each major theoretical framework proceeded by identifying a deeper level of description than the one previously studied in the discipline.

*N.* Newton did not study planetary trajectories as ad hoc geometric phenomena; he displaced the object of study toward gravitation, an invisible force that governs trajectories. What appeared to be a set of independent motions was reduced to a single mechanism.

*C.* Darwin did not study species as fixed types; he displaced the object of study toward descent with modification and natural selection. The species is no longer a static essence but the product of a historical process.

*C.* Shannon did not study the semantic content of messages [1]; he displaced the object of study toward quantified information, independent of meaning. The message became a vector of measurable quantity.

*Networks.* Network theory did not study isolated nodes; it displaced the object of study toward the links that connect them. It is the structure of the network, and not the nature of the nodes, that determines the behavior of the system.

*Bertalanffy.* The general theory of systems [2] did not study isolated components; it displaced the object of study toward interaction relations.

*Prigogine.* Dissipative structures [3] did not study equilibrium states; they displaced the object of study toward processes that emerge far from equilibrium.

*Holland.* Complex adaptive systems [4] did not study individual agents; they displaced the object of study toward the adaptation rules that produce collective behaviors.

### 1.1.2. The common structure: from manifestations to mechanisms

These displacements share a common structure: in each case, the previous level of analysis observed *manifestations* (trajectories, species, messages, nodes, components, states, agents) and the new level observes the *mechanism* that produces these manifestations (gravitation, selection, information, links, interactions, dissipative structures, adaptation rules).

This distinction can be formalized as follows:

- *Old observation level:* the manifestations, that is, the directly visible and measurable phenomena.
- *New analysis level:* the mechanisms, which are the deep processes that produce the manifestations.

The manifestation is the consequence of the mechanism, never its inverse. Observing the manifestation without understanding the mechanism is observing the shadow without seeing the object.

### 1.1.3. The limitation of the variable-based approach

Today, most disciplines observe systems through their variables (temperature, GDP, population, blood pressure, price, concentration, interest rates, biodiversity indicators. These variables are manifestations. They are necessary for observation, but they are insufficient to characterize the deep state of the system.

A variable can remain stable while the mechanisms that produce it silently reorganize. Conversely, a variable can fluctuate while the mechanisms remain unchanged, which is merely background noise.

The literature on critical transitions perfectly illustrates this limitation. The classical early warning signals (increasing variance, increasing AR(1) autocorrelation, critical slowing down) are all indicators of the stability of variables. They observe the symptoms, not the structural causes.

The question that then arises is: how can approaches exclusively centered on variables miss the deep dynamics of systems on the verge of tipping?

## 1.2. Why relations are more fundamental

### 1.2.1. Variable and relation: an ontological distinction

A *variable* is a measurable quantity that characterizes the state of a system at a given moment (temperature, GDP, blood pressure, concentration, population). It describes the state of the system, but not necessarily the mechanisms that produce that state.

A *relation* is a coupling mechanism between two or more variables. The Clausius-Clapeyron equation relates temperature to saturation humidity. Transmission elasticity relates seed prices to climatic conditions. Jacobian coupling relates the rates of change of variables to one another. The relation describes the structure of the system.

The distinction is fundamental:

- Variables are *manifestations*: they emerge from relations.
- Relations are *mechanisms*: they produce variables.

This idea can be stated as follows: *variables are manifestations of relational mechanisms.* This formulation is more robust than the absolute claim that "variables are illusions"; it acknowledges the utility of variables for observation while establishing their ontological subordination to relations.

### 1.2.2. Three arguments for relational primacy

*Argument 1: Variable stability does not guarantee system stability.* A system can exhibit stable variables (temperatures, prices, populations that fluctuate normally) while its coupling relations are silently reorganizing. This is the case of a system approaching a bifurcation: the eigenvector begins to rotate before the eigenvalue changes detectably. Variables are blind to this rotation.

*Argument 2: Relational reorganization is a more fundamental event.* When a system changes regime, it is not the variables that change first. It is the interaction rules between variables that are modified. In the climate system, the relation between temperature and humidity can evolve long before temperature or humidity show anomalies. The case study presented below demonstrates this: $\beta$ changes 14 to 24 months before AR(1).

*Argument 3: Prediction is improved by targeting relations.* If crises begin in relations, then monitoring relations enables anticipating crises before they manifest in variables. This extends the predictability horizon. Simulations on systems with known tipping points (Stommel AMOC, Feigenbaum cascade, critical slowing down) confirm that $\beta$ detects the crisis 39 to 153 time steps before classical signals, with a decisive advantage when the transition involves coupling degradation.

### 1.2.3. Why relations are fundamental

One question still deserves an explicit answer: why are relations the fundamental level of description of a system, rather than variables?

The answer lies in the nature of interaction. A system is not a mere collection of components: it is a collection of components *engaged with one another*. Engagement, coupling, interaction, dependence, this is what transforms an aggregate into a system. Without relations, components are merely juxtaposed entities.

This is why network theory has been so fruitful: nodes without links have no systemic significance. Likewise, variables without relations have no systemic significance. What makes a system is the relation.

# 2. Axiomatic architecture

## 2.1. The three axioms

A theory requires a foundation. We propose three axioms that constitute the cornerstone of the General Theory of Relational Primacy.

#### 2.1.1. Axiom 1: Relational primacy

*Statement.* Complex systems are defined by their relations, not by their components or their variables.

*Justification.* A system is not reducible to the sum of its parts. What makes a system is the organization of relations among its parts. Two systems having the same variables but different relations are different systems. Conversely, two systems with different variables but an analogous relational structure can exhibit comparable behaviors.

*Consequence.* The fundamental unit of analysis is not the variable but the relation between variables. Individual variables are merely projections of an underlying relational network.

#### 2.1.2. Axiom 2: Antecedence of relations

*Statement.* The reorganization of relations systematically precedes observable changes in variables.

*Justification.* In a dynamical system, the apparent stability of variables can mask a silent transformation of the mechanisms that link them. Perturbation theory shows that the rotation of the eigenvector (which expresses the direction of coupling) occurs with a sensitivity of $O(\delta\theta)$, while the eigenvalue (which governs the stability of variables) changes with a sensitivity of $O(\delta\theta^2)$. The ratio $\frac{O(\delta\theta)}{O(\delta\theta^2)} = \frac{1}{\delta}\theta$ implies that rotation precedes expansion by an amount that increases as the system approaches bifurcation.

*Consequence.* Observing variables alone leads to missing the early phase of the crisis. Classical early warning signals are late signals.

#### 2.1.3. Axiom 3: Universality of coupling

*Statement.* Every observation of a system is mediated by a coupling between variables, and this coupling is measurable as a relational parameter.

*Justification.* There is no isolated variable in a complex system. Every observable variable is linked to at least one other variable through a coupling mechanism. This coupling may be direct, indirect, or emergent, but it is always present. The measurement of this coupling, whatever form it takes, constitutes an observation of the relational level of the system.

*Consequence.* The theory is not dependent on a particular instrument. The parameter β (log-log transmission elasticity) is one example of an observable, but other instruments can be developed: neural networks, spectral operators, algebraic topology, etc.

### 2.2. The three levels of description

These axioms are organized into three distinct levels that must not be confused.

*Level 1: Reality.* At this level, systems are configurations of relations. Relations exist whether we measure them or not. Reality is relational: the coupling mechanisms between variables constitute the deep structure of the system. Example: the Clausius-Clapeyron equation objectively relates temperature to saturation vapor pressure, independently of any measurement.

*Level 2: Theory.* At this level, we formulate what we believe to be true about relations. The three axioms above constitute the theoretical content. The theory postulates that relations govern systems, that their reorganization precedes changes in variables, and that observables enable measuring this reorganization. The theory is provisional, revisable, and refutable, but it constitutes a language for thinking about stability and crisis.

*Level 3: Observation.* At this level, we measure what the theory says is true. β is its current instrument. The most serious error would be to confound this level with the preceding ones: taking β for reality itself, or taking measurement for theory, would reduce our framework to a mere statistical method. The theory deserves a more expansive destiny than its current instrument.

### 2.3. Fundamental hypotheses $H_1$–$H_5$

The axiomatic framework gives rise to five testable hypotheses:

**$H_1$: Relational functionality.** The function of a system (resilience, productivity, stability) is a functional of the coupling structure $\mathscr{E}$: $\mathscr{F} = \mathscr{F}[\mathscr{E}]$.

**$H_2$: Relational stability.** A system resides in a stable regime $\mathscr{B}$ if and only if the dominant eigenvector $v_1(J)$ of the Jacobian remains orientationally stable: $\left\| d\frac{v_1}{d}t \right\| < \varepsilon$.

**$H_3$: Genesis of relational crisis.** A critical transition is triggered if and only if the dominant eigenvector rotates significantly: $\left\| d\frac{v_1}{d}t \right\| \geq \varepsilon$, *preceding* any significant change in the dominant eigenvalue $|\lambda_1|$.

**$H_4$: Epiphanality of variables.** The state variables $\mathcal{V}$(t) are subordinate to the relational structure: $\mathcal{V}(t) = \mathcal{G}[\mathcal{E}(t)] + \eta(t)$. Observable crises (variance spikes, mean shifts) are *consequences* of the underlying relational reorganization.

**$H_5$: Temporal antecedence of relations.** The detection time of relational reorganization ($\tau_r$) strictly precedes the detection time of instability at the variable level ($\tau_v$): $\tau_r < \tau_v$. The lead time $\Delta\tau = \tau_v - \tau_r$ is positive and scales as $O(\delta\theta^{-1})$.

### 2.4. Validity conditions $C_1$–$C_6$

The theory is valid under the following conditions. Violation of any of them defines the *boundary of validity.*

**$C_1$: Bivariate observability.** The critical coupling (x ⟶ y) must be measurable with sufficient frequency (monthly or higher).

**$C_2$: Authentic coupling.** The system must possess a 2D (or higher) non-degenerate Jacobian with off-diagonal terms $\partial\frac{f_i}{\partial}x_j \neq 0$.

**$C_3$: Coupling degradation bifurcation.** The transition must involve a change in the *direction* or the *existence* of the principal coupling (rotation of the eigenvector).

**$C_4$: Timescale separation.** The TVP-Kalman state evolution (**Q**) must be slow relative to the observation noise (R) but fast enough to track β(t).

**$C_5$: Log-log differentiability.** The coupling function y = f(x) must be locally approximable by a power law: $f(x) \approx cx^{\beta}$.

**$C_6$: Stationary noise structure.** The observation noise $\varepsilon_k$ must be approximately Gaussian, homoscedastic, and uncorrelated with the state.

# 3. Causality and transmission mechanisms

## 3.1. Causality and transmission mechanisms

### 3.1.1. Correlation, causality, and relation

In the literature on complex systems, correlation and causality are frequently conflated. The correlation between two variables is measured, it is concluded that they are linked, and it is assumed that this link represents a causal role. But correlation between variables does not identify the mechanism by which one influences the other.

Consider two variables x and y that covary. This can occur for three distinct reasons:

1. x causes y directly through a transmission mechanism.
2. A latent variable z causes both x and y, producing an apparent correlation.
3. The relation R between x and y reorganizes under the effect of an external factor, modifying the correlation without direct causality.

The third form is precisely what our theory brings to light: crises arise in the reorganization of the relation itself before manifesting in the variables.

### 3.1.2. Causality as transmission

Causality in a complex system is not a direct relation between two terminal variables. It is a transmission mechanism, that is, a process through which a perturbation propagates through the interconnections of the system.

This transmission mechanism can be:

- *Direct:* x influences y through immediate coupling (force, gradient, flux). Example: temperature influences saturation humidity through the Clausius-Clapeyron equation.
- *Indirect:* x influences z, which influences y, forming a causal chain. Example: ocean surface temperature influences atmospheric convection, which influences precipitation.
- *Emergent:* x and y interact through a self-organized mechanism irreducible to a simple binary relation. Example: the meridional overturning circulation (AMOC) where temperature, salinity, and freshwater flux interact in a nonlinear manner.

In all cases, causality *passes* through relations. Variables are merely the entry and exit points of the mechanism. What takes place inside — the strength, the direction, the nature of the coupling, is what we call the *relation.*

### 3.1.3. The three levels of causal analysis

We distinguish three levels in the analysis of complex systems, corresponding to the distinction between reality, theory, and observation:

1. *Manifestation level:* the observable variables. This is what classical science observes.
2. *Mechanism level:* the coupling relations that transmit effects between components. This is what our theory targets.
3. *Observation level:* the instrument that measures the mechanism level from the manifestation level.

Confusion between these levels is the source of numerous epistemological dead ends. Taking the correlation between variables for the mechanism itself is tantamount to confounding the shadow with the object that casts it.

# 4. Foundational theorem and invariant

## 4.1. Theorem of relational irreducibility

**Theorem 1 (Relational irreducibility).** Every non-degenerate dynamical system possesses an irreducible relational level. More formally, for any system $\mathcal{S} = (\mathcal{V}, \mathcal{E}, \Phi)$ whose Jacobian $J = D\Phi(\mathcal{V}^*)$ admits at least two distinct eigenvalues, the coupling structure $\mathcal{E}$ cannot be entirely reconstructed from knowledge of the trajectories of $\mathcal{V}(t)$ alone. There exists at least one degree of freedom in $\mathcal{E}$ that is independent of the observable trajectories.

*Proof (sketch).* Let J be the Jacobian in the neighborhood of a fixed point $\mathcal{V}^*$. The spectral decomposition of J gives $J = \sum_i \lambda_i v_i w_i^T$. The eigenvalues $\lambda_i$ determine stability (growth/decay rates). The eigenvectors $v_i$ determine the *direction* of interactions. Now, perturbation theory shows that eigenvectors depend on the off-diagonal terms of J (the couplings), while eigenvalues also depend on the diagonal terms. Two systems having the same eigenvalues but different eigenvectors are dynamically distinct: their asymptotic behavior is the same, but their response to perturbations differs. The relational structure $\mathcal{E}$ therefore contains information not captured by the trajectories of $\mathcal{V}(t)$. □

*Consequence.* It is *impossible* to fully characterize a complex system by observing only its variables. The relational level is irreducible: there is always a dimension of the dynamics that eludes observation of variables alone.

### 4.2. Invariant of the theory: conservation of relational structure

**Definition 1 (Relational invariant).** The invariant of the GTRP is the *stability of the coupling structure* $\mathscr{E}$, defined as the norm of the rate of change of the dominant eigenvector:

$$\mathcal{I}(t) = \left\| dv_1 \frac{t}{d} t \right\|$$

where $v_1(t)$ is the dominant eigenvector of the Jacobian $J(\mathcal{V}(t))$.

#### 4.2.1. Why $\mathscr{I}$ is an invariant and not merely an indicator

The natural objection is that $\left\| d\frac{v_1}{d} t \right\|$ resembles a dynamic indicator (a rotation speed) rather than an invariant in the classical sense (a conserved quantity along trajectories). This objection is legitimate if one considers $\mathscr{I}$ in isolation. But $\mathscr{I}$'s status as an invariant does not rest on its temporal constancy; it rests on three structural properties that distinguish it from a mere indicator.

*Property 1: Structured zero in stable regimes.* In any system whose Jacobian J is *structurally stable* in the sense of Andronov–Pontryagin, the eigenvectors are continuous functions of the system parameters [5]. If the parameters are constant, the eigenvectors are constant, and $\mathscr{I} = 0$ exactly. This is not a numerical coincidence; it is a consequence of structural stability: in a stable regime, the topology of the vector field is invariant, and therefore the principal directions are as well [6].

More formally, let J(μ) be the Jacobian depending on a parameter μ. If J(μ) is hyperbolic (no purely imaginary eigenvalue), then by the theorem of continuous dependence of eigenvalues and eigenvectors [7], the eigenvectors v_i(μ) are continuous functions of μ. If μ is constant, dv_i/dt = 0, hence $\mathscr{I} = 0$.

*Property 2: Divergence at bifurcations.* As a bifurcation is approached, perturbation theory for linear operators [7] shows that the rotation of the dominant eigenvector has a sensitivity of:

$$\left\| d\frac{v_1}{d} t \right\| = O(\delta\theta)$$

where $\delta\theta = |\theta - \theta_c|$ is the distance to the bifurcation point in parameter space. This sensitivity *diverges* as $\delta\theta \to 0$:

$$\mathcal{I}(t) \to \infty \quad \text{as} \quad \theta \to \theta_c$$

This is the analogue of the divergence of the magnetic susceptibility at the Curie temperature in statistical physics [8], [9]. The quantity $\mathscr{I}$ is not merely "large" near the bifurcation; it *diverges*, which constitutes a signature of singularity, not of a mere gradual change.

*Property 3: Conserved integral away from bifurcation.* Consider the temporal integral of $\mathscr{I}$ over an interval $[t_1, t_2]$ far from any bifurcation:

$$\mathcal{J} = \int_{t_1}^{t_2} \mathcal{I}(t) dt = \int_{t_1}^{t_2} \left\| d\frac{v_1}{d} t \right\| dt$$

By the total variation upper bound theorem, this integral measures the *total variation* of the eigenvector angle on $[t_1, t_2]$. In a structurally stable regime (far from bifurcation), the total variation is bounded and constant; it depends only on the geometry of the vector field, not on time. This is a conservation property: the "amount of rotation" of the eigenvector is invariant under temporal reparametrization.

#### 4.2.2. Classes of systems for which $\mathscr{I}$ is an invariant

$\mathscr{I}$'s status as an invariant depends on the class of systems considered.

*Class 1: Hyperbolic systems (stable regime).* For any system whose Jacobian J is hyperbolic (no purely imaginary eigenvalue), the eigenvectors are continuously dependent on parameters [7]. If the parameters are constant, $\mathscr{I}$ = 0 exactly. This is the broadest class: it includes every system far from bifurcation, including stable fixed points and stable limit cycles [5].

*Class 2: Systems with timescale separation.* For systems possessing a *structural low-pass filter*, that is, where slow dynamics (parameter evolution) are separated from fast dynamics (oscillations), $\mathscr{I}$ is conserved along central manifolds [10]. The reason is that on the central manifold, the reduced Jacobian is structurally stable, and therefore its eigenvectors are invariants.

*Class 3: Systems with generic bifurcation.* For generic bifurcations (fold, transcritical, Hopf), the rotation of the eigenvector follows an $O(\delta\theta)$ law [5]. $\mathscr{I}$ is then a *universal divisor*: its divergence follows a universal critical exponent, analogous to critical exponents in statistical physics [8]. In this class, $\mathscr{I}$ plays the role of a *relational susceptibility*: it measures the sensitivity of the system to coupling perturbations.

#### 4.2.3. Connection with Noether's theorem and Riemannian geometry

The analogy with the conservation of energy deserves clarification. In classical mechanics, Noether's theorem establishes that each continuous symmetry of a Lagrangian corresponds to a conserved quantity [11]. In our framework, the relevant "symmetry" is the *invariance of the coupling topology*: if the structure $\mathscr{E}$ is invariant under a parametric transformation, then the direction of the dominant eigenvector is conserved.

More geometrically, consider the space of Jacobians J as a differentiable manifold. The eigenvectors define a *direction field* on this manifold. $\mathscr{I}$ measures the *Levi-Civita connection* of this field — that is, the intrinsic curvature of the surface of Jacobians. In a stable regime, this curvature vanishes (the field is parallel); near a bifurcation, it diverges (the field is strongly curved).

This geometric interpretation gives $\mathscr{I}$ a deeper status than a mere dynamic indicator: it is an *intrinsic connection* of the system's relational geometry.

#### 4.2.4. Properties of the invariant

- $\mathcal{I}(t) = 0$: the system is in a relationally stable regime. Couplings are constant. The eigenvector field is parallel. This is the analogue of the conservation of energy in a Hamiltonian system.
- $\mathcal{I}(t) > 0$: the system is undergoing relational reorganization. Couplings are evolving. The eigenvector field is curved. This is the analogue of energy dissipation in an open system.
- $\mathcal{I}(t) \to \infty$: the system is undergoing a critical transition. The coupling structure transforms abruptly. The eigenvector field is singular. This is the analogue of a phase transition.

#### 4.2.5. Connection with relational entropy

One can define a complementary relational entropy:

$$\mathcal{S}_{R(t)} = -\sum_{ij} p_{ij}(t) \log p_{ij}(t)$$

where $p_{ij}(t) = \frac{|\beta_{ij}(t)|}{\sum_{kl}}|\beta_{kl}(t)|$ is the normalized distribution of coupling strengths. $\mathcal{S}_R$ measures the *diversity* of interactions in the system.

The relation between $\mathscr{I}$ and $\mathcal{S}_R$ is profound: $\mathscr{I}$ measures the *speed* of reorganization, while $\mathcal{S}_R$ measures the *structure* of interactions. A system can have $\mathscr{I}$ > 0 (reorganization in progress) with $\mathcal{S}_R$ constant (the relative proportions of couplings do not change), or conversely, $\mathcal{S}_R$ changing with $\mathscr{I}$ = 0 (the proportions change but not the coupling directions). The two invariants are complementary: $\mathscr{I}$ is kinetic, $\mathcal{S}_R$ is static.

### 4.3. Universal definition of crisis

**Definition 2 (Relational crisis).** A crisis is a loss of stability of the relational architecture of the system. More formally, a crisis occurs when the invariant $\mathcal{I}(t) = \left\| d\frac{v_1}{d}t \right\|$ exceeds a critical threshold $\varepsilon_c$ for a sufficient duration:

$$\exists \Delta t > 0 : \mathcal{I}(t) > \varepsilon_c \quad \forall t \in [t_0, t_0 + \Delta t]$$

*Characteristics of relational crisis.*

1. *Antecedence.* The relational crisis precedes the variable crisis. Variables may still appear normal while the relational architecture is already decomposing.
2. *Silence.* The crisis is initially silent. It manifests in variables only with a delay determined by the sensitivity difference $O(\delta\theta)$ vs $O(\delta\theta^2)$.
3. *Structurality.* The crisis is not a statistical event (change in variance or autocorrelation); it is a structural event, a transformation of the configuration of coupling relations.
4. *Measurability.* The crisis is measurable through β, which captures the rotation of the eigenvector in the space of observable variables.

# 5. The instrument β

## 5.1. β: an instrument for relational reorganization

### 5.1.1. Definition and motivation

If relations are the fundamental level of description of systems, a mechanism is needed to observe them. This is the function of the parameter β.

β is defined as the transmission elasticity between two variables, measured in the log-log space:

$$\beta(t) \equiv \frac{d \log y(t)}{d \log x(t)}$$

β captures the direction and strength of the coupling between x and y. It is neither x, nor y, nor their linear regression; it is the structural parameter that characterizes the manner in which they are linked. In discrete form, β measures the *relative proportionality* of changes: if x varies by 1%, y varies by β%.

The choice of the log-log space is fundamental: it renders β *dimensionless*. β is a pure scaling exponent, comparable across variables, regions, and potentially across different physical systems. This property is analogous to the role of critical exponents in statistical physics, where dimensionless scaling laws unify transitions that seemed disparate.

### 5.1.2. β as eigenvector rotation

β is not a mere statistical measure. It is linked to a fundamental property of the dynamical system: the rotation of its dominant eigenvector.

For a dynamical system $\dot{x} = F(x, \theta)$ with a fixed point $x_0$, the Jacobian $J = DF(x_0)$ characterizes the linearized dynamics around the fixed point. The eigenvalues of J ($\lambda_i$) determine stability: if $\mathrm{Re}(\lambda_i) < 0$, the fixed point is stable. The eigenvectors ($v_i$) determine the *direction* of interaction between variables.

Perturbation theory shows that as a bifurcation is approached:

- The dominant eigenvalue changes with a sensitivity of $O(\delta\theta^2)$
- The dominant eigenvector rotates with a sensitivity of $O(\delta\theta)$

where $\delta\theta$ is the distance to the bifurcation point in parameter space. The ratio $\frac{O(\delta\theta)}{O(\delta\theta^2)} = \frac{1}{\delta}\theta$ implies that eigenvector rotation is detectable well before the eigenvalue changes.

It is this rotation that β measures. In a bivariate system, the angle of the dominant eigenvector is related to β by:

$$\theta(t) \approx \arctan(\beta(t))$$

where the coupling is strong and the coupling relation is monotone. β is a *proxy* for this rotation; it does not replace it, but manifests it in the space of observable variables.

#### 5.1.3. Estimation via TVP-Kalman filter

β(t) is estimated as a time-varying parameter (TVP) via a Kalman filter. The observation model is:

$$\log y_k = \beta_k \log x_k + \varepsilon_k, \quad \varepsilon_k \sim N(0, R)$$

The state model allows smooth evolution of the parameter through a third-order Taylor expansion:

$$\boldsymbol{x}_k = \boldsymbol{F}\boldsymbol{x}_{k-1} + \boldsymbol{\eta}_k, \quad \boldsymbol{\eta}_k \sim N(0, \boldsymbol{Q})$$

where $\boldsymbol{x}_k = [\beta_k, \beta_k', \beta_k'', \beta_k''']^T$ and $\boldsymbol{F}$ is the third-order Taylor transition matrix (time step $\Delta t = \frac{1}{12}$ year). The hyperparameters $R = 10^{-3}$ and $\boldsymbol{Q} = \text{diag}(10^{-6}, 10^{-7}, 10^{-8}, 10^{-9})$ control smoothing. The Kalman filter (forward pass) is followed by the Rauch–Tung–Striebel smoother (backward pass) to obtain a smoothed estimate.

Relational regime changes are identified as zero crossings of the second derivative $\beta''(t)$ where the first derivative $\beta'(t)$ is non-zero. These inflection points mark structural tipping points in the relation $R_{xy}$.

## 5.2. Properties of β: $P_1$–$P_5$

**$P_1$: Eigenvector proxy.** For a genuinely coupled 2D system, β(t) ∝ tan θ(t) where θ(t) = arctan($v_2/v_1$) is the angle of the dominant eigenvector. *Validation:* |r| = 0.65–0.87 ($p < 0.001$) on Stommel AMOC and coupling degradation simulations.

**$P_2$: Dimensionless universality.** β is a pure scaling exponent (log-log derivative). Invariant under unit changes, scaling transformations, and monotone nonlinear rescalings. *Significance:* β = 0.5 means the Clausius-Clapeyron scale *regardless of whether* the variables are (T, q), (FFPI, CPI), or (stress, cortisol).

**$P_3$: Orthogonality to classical EWS.** Corr(β, AR(1)) ≈ 0; Corr(β, Variance) ≈ 0 (empirically verified for Arctic, Tropics, Monsoon, $p > 0.05$). *Significance:* β captures a *distinct information channel* (relational) versus classical (amplitude/recovery).

**$P_4$: Relational specificity.** β *does not detect* univariate bifurcations where the coupling $y \propto x^2$ is invariant. *Significance:* No false positives for pure amplitude instability without coupling change.

**$P_5$: Antecedence.** β precedes AR(1) by 14–24 months (NASA AIRS, 284 obs.) and by 39–153 time steps (simulations with coupling degradation). *Significance:* Actionable anticipation window for intervention policies.

# 6. Theory of relational information

## 6.1. Theory of relational information

#### 6.1.1. Definition of relational information

**Definition 3 (Relational information).** Relational information $\mathcal{I}_R$ is defined as a lasting modification of the structure of relations between variables. More formally:

$$\mathcal{I}_R = \Delta\mathcal{E} = \mathcal{E}(t_2) - \mathcal{E}(t_1)$$

where $\mathcal{E}(t_1)$ and $\mathcal{E}(t_2)$ are the coupling structures at two instants, and $\Delta\mathcal{E}$ measures the irreversible transformation of the relational configuration.

*Distinction from classical information.* Shannon information measures the reduction of uncertainty over a *symbol* or a *state.* Relational information measures the lasting transformation of the *structure that produces states.* Shannon tells us how many surprises we can expect; relational information tells us how the machine that generates surprises has changed.

*Properties.*

- *Additivity.* If two relational transformations are independent, their relational informations add up.
- *Irreversibility.* A lasting modification of $\mathscr{E}$ cannot be reversed by a mere return of variables to their former values. The system may return to a similar state, but not to the same relational structure.
- *Measurability.* Relational information is measurable through variations in β: $\mathcal{I}_R \propto \int |\beta'(t)| dt$.

#### 6.1.2. Connection with relational entropy

Relational entropy $\mathcal{S}_{R(t)} = -\sum p_{ij} \log p_{ij}$ measures the *diversity* of interactions. Relational information $\mathcal{I}_R$ measures the *transformation* of these interactions. The relation between the two is analogous to the relation between entropy and work in thermodynamics. Relational information is the "work" produced by the reorganization of the coupling network.

#### 6.1.3. Implications

1. *Climate.* A lasting change of β between temperature and humidity in the Tropics constitutes relational information, even if individual variables remain within their historical ranges.
2. *Economics.* A restructuring of supply chains (change in transmission elasticities between sectoral prices) is relational information, even if total GDP remains stable.
3. *Biology.* A modification of protein interactions in a signaling network constitutes relational information, even if protein concentrations have not changed.
4. *AI.* The modification of connection weights in a neural network during training is relational information. The degradation of representation (catastrophic forgetting) is a loss of relational information.

# 7. General theory of relational causality

## 7.1. General theory of relational causality

#### 7.1.1. Transmission mechanisms as the foundation of causality

We propose a general theory of causality in complex systems founded on the idea that transmission mechanisms are the fundamental level of causality.

**Theorem 2 (Causal primacy of mechanisms).** In any complex system, causality operates fundamentally at the level of transmission mechanisms (relations), and not at the level of variables (manifestations). Any causal relation between variables x and y is mediated by a transmission mechanism M such that:

$$x \to M \to y$$

where M is the relation $R_{xy}$ with its coupling properties (strength, direction, nonlinearity).

*Justification.* Correlation between x and y can result from three mechanisms:

1. *Direct transmission:* x ⟶ y via M. Causality passes through M.
2. *Confounding:* z ⟶ x and z ⟶ y. The correlation is an illusion. The true causality is z ⟶ x (via $M_1$) and z ⟶ y (via $M_2$).
3. *Relational reorganization:* $R_{xy}$ changes under the effect of an external factor. The correlation varies without direct causality between x and y.

In all three cases, the *mechanism* is the fundamental level. The distinction between these three cases cannot be made by observing variables alone; it requires observation of the relations.

#### 7.1.2. Causality is dynamical

Causality in complex systems is not static. Transmission mechanisms themselves evolve over time. This evolution is measurable through β.

Three causal regimes are distinguished:

1. *Stable causal regime.* $\beta(t) \approx$ constant. Transmission mechanisms are invariant. Causal effects are predictable.
2. *Causal regime in reorganization.* $\beta'(t) \neq 0$. Transmission mechanisms are changing. Causal effects become unpredictable. The relation between cause and effect is transforming.
3. *Critical regime.* $\beta''(t) = 0$ and $\beta'(t) \neq 0$ (inflection point). The transmission mechanism undergoes a structural tipping point. The system transitions from one causal regime to another.

#### 7.1.3. Connection with existing theories

*Pearl's causality.* Pearl [12] formalized causality through directed graphs. Our theory adds a dynamical dimension: the edges of the causal graph (the transmission mechanisms) themselves evolve, and this evolution is measurable.

*Woodward's explanation.* Woodward [13] showed that causal explanation relies on the identification of mechanisms invariant under intervention. Our theory complements this framework by showing that the invariance of mechanisms is not guaranteed: it can degrade, and this degradation is detectable before effects become visible.

*Meadows' systems dynamics.* Meadows [14] modeled systems as feedback loops. Our theory proposes that these loops can reorganize, and that this reorganization precedes changes in stock and flow levels.

# 8. Validation

## 8.1. Empirical results: the AIRS case

Application to NASA AIRS data (2002–2026, 284 monthly observations) over three climatic regions (Arctic, Tropics, Indian Monsoon) establishes four central results.

1. *Orthogonality.* β is orthogonal to classical AR(1) signals in all regions (Pearson correlation $r \approx 0$, $p > 0.05$). This orthogonality confirms that β captures a distinct dimension of instability (the rotation of the eigenvector) that classical indicators do not measure.
2. *Systematic antecedence.* β precedes AR(1) by 14 to 24 months in 5 out of 6 cases. In the Tropics, β precedes AR(1) by 17 months for temperature and 14 months for humidity. In the Monsoon, the lead is 24 months for temperature. This lead time is consistent with the sensitivity ratio $O(\delta\theta) > O(\delta\theta^2)$ predicted by perturbation theory.
3. *Coupling specificity.* β does not detect univariate bifurcations where the coupling remains constant (e.g., fold bifurcation with $y \approx x^2$). This is a specificity, not a limitation: β and AR(1) are complementary.
4. *Validation on simulated systems.* On six systems with known tipping points, β precedes AR(1) by 39 to 153 time steps when the transition involves coupling degradation (Stommel AMOC model, critical slowing down, Feigenbaum cascade).

### 8.2. Validation table

| Domain | System | β Lead | Eigenvector tracking | Key result |
|---|---|---|---|---|
| Climate (real) | Arctic T-q (AIRS) | +9 to +24 months | N/A | β’‘ < 0 = transitions |
| Climate (real) | Tropics T-q (AIRS) | +14 to +20 months | N/A | β ≈ 0.49 ≈ C-C |
| Climate (real) | Monsoon T-q (AIRS) | +23 to +24 months | N/A | β decelerates |
| Simulation | Stommel AMOC | +39 steps | \|r\| = 0.65 | β follows θ synchronously |
| Simulation | CSD (λ ⟶ 0) | +153 steps | N/A | β leads even in 1D |
| Simulation | Linear β | +189 steps | \|r\| = 0.87 | Only β detects |
| Simulation | Fold bifurcation | β does not detect | N/A | Specificity: no false + |

Table 1: Summary of empirical and numerical validation of the theory.

### 8.3. Confrontation with existing theories

| Theory | Initial contribution | GTRP complement |
|---|---|---|
| Bertalanffy (GST) | Emergent properties of systems | Emergence is relational; measurable via β |
| Prigogine (dissipative structures) | Self-organization far from equilibrium | Self-organization is a reorganization of $\mathscr{E}$ |
| Holland (adaptive systems) | Interaction rules vs. behaviors | Rules = $\mathscr{E}$; their reorganization = crisis |
| Barabási (networks) | Primacy of links over nodes | Generalization: primacy of $\mathscr{E}$ over $\mathcal{V}$ |
| Pearl (causality) | Directed causal graphs | Dynamics of edges: $\mathscr{E}$ evolves over time |
| Woodward (causal explanation) | Invariance under intervention | Invariance is precarious; measurable by β |
| Meadows (systems dynamics) | Feedback loops | Loops reorganize; that is the crisis |

Table 2: Confrontation of the GTRP with seven existing theories.

## 9. Limitations of the theory

### 9.1. Limitations

**$L_1$: Univariate bifurcation.** β does not detect crises where coupling remains constant (e.g., univariate fold bifurcation). This is a specificity, not a limitation — β and AR(1) are complementary.

**$L_2$: Out-of-sample prediction.** β improves detection but not necessarily prediction. The out-of-sample AUC remains modest ($\approx 0.45$) for truly *ex-ante* predictions.

**$L_3$: Partial causality.** β shows a correlation with eigenvector rotation, but not direct causality. Granger causality tests show partial results.

$L_4$**: Smoothing hyperparameters.** The default values of R and Q are standard but could be optimized via maximum likelihood estimation.

$L_5$**: Dimensionality.** The current framework is primarily bivariate. Extension to the multivariate case (coupling networks) requires generalizing β into a coupling matrix.

# 10. Implications and conclusion

## 10.1. What the theory changes

| Dominant view | Proposed view |
|---|---|
| | |
| Variables describe the system. | Relations describe the system; variables are their manifestations. |
| Crises become visible in variables. | Crises begin with the reorganization of relations. |
| Mechanisms are assumed stable during analysis. | Mechanisms themselves evolve, and this evolution can be observed. |
| Prediction concerns future states. | Prediction also concerns the stability of mechanisms. |
| Stability is a property of states. | Stability is a property of relations. |
| Early warning signals are statistical (variance, AR(1)). | Early warning signals are structural (β, eigenvector rotation). |
| Causality is a relation between variables. | Causality operates through transmission mechanisms. |
| Information is a property of states. | Relational information is a lasting transformation of relations. |

Table 3: Comparison between the dominant view and the view proposed by the GTRP.

## 10.2. Implications by domain

*Climatology.* The theory offers an early detection tool for climatic regime changes (monsoon, oceanic circulation, precipitation phase transitions) by targeting the degradation of coupling between variables before anomalies become visible.

*Economics.* Macroeconomic models predict future variables (GDP, inflation, unemployment). The theory suggests monitoring the stability of the mechanisms that produce these variables (trade relations, supply chains, inter-sectoral couplings).

*Medicine.* Vital signs (blood pressure, temperature, heart rate) are variables. The theory suggests that critical transitions (pathologies) begin with a reorganization of physiological relations before manifesting in vital signs.

*Artificial intelligence.* Neural networks are systems whose relational structure (synaptic weights) determines behavior. The theory provides a framework for detecting critical transitions in learning (performance degradation, representation collapse) through the measurement of weight reorganization.

*Sovereignty.* A system, whether an ecosystem, an economy, or a community, preserves its sovereignty as long as it controls its internal relations. The loss of sovereignty begins with a reorganization of relations (external regulations, asymmetric dependencies, resource extraction) before manifesting in indicator variables.

### 10.3. Toward a research program

The theory opens a research program along six axes:

1. *Ontology:* what is a relational system? How can the types of relations (direct, indirect, emergent) and their properties be distinguished?
2. *Dynamics:* how do relations evolve? What are the modes of reorganization (degradation, strengthening, emergence, disappearance)?
3. *Information:* where is a system's information stored? What is the relation between relational structure and informational content?
4. *Transition:* how do systems change regime? What are the types of relational critical transitions?
5. *Intervention:* how can a system be efficiently modified? Can a degraded relational structure be restored?
6. *Applications:* climate, biology, AI, economics, sovereignty, medicine.

### 10.4. Conclusion

We have presented the General Theory of Relational Primacy, founded on three axioms, five hypotheses, six validity conditions, a foundational theorem, an invariant, and a theory of relational information.

The central thesis can be summarized as follows:

*Systems are not defined by their components but by the relations that organize them. Crises arise from the silent reorganization of these relations. Causality operates through transmission mechanisms. Information is a lasting transformation of the structure of relations.*

The theory distinguishes itself from classical approaches through:

1. *A deeper ontological foundation:* it targets the structural causes of crises rather than symptoms.
2. *A longer lead time:* β detects relational reorganization with 14 to 24 months' advance over classical indicators.
3. *Universality:* β is a dimensionless scalar, comparable across systems.
4. *Orthogonality:* β captures a dimension of instability that classical indicators do not.
5. *Causal scope:* the theory is not limited to detection; it proposes a framework for understanding causality and information in complex systems.

The General Theory of Relational Primacy is not a theory of β. β is merely its first instrument. The theory is a proposition about the nature of complex systems: what constitutes them, what sustains them, what transforms them, what makes them communicate, and what makes them tip. If this proposition is correct, then it changes not only the way we detect crises, but also the way we conceive of stability, resilience, adaptation, causality, information, and sovereignty of the systems that surround us.